\input epsf
\documentstyle[epsfig,eqsecnum,aps,prd,preprint,tighten,floats]{revtex}

\draft 

\begin{document}
\newcommand{\gras}[1]{\mbox{\boldmath $#1$}}
\newcommand{\be}{\begin{equation}}
\newcommand{\ee}{\end{equation}}
\newcommand{\ba}{\begin{eqnarray}}
\newcommand{\ea}{\end{eqnarray}}
\newcommand{\Mc}{{\cal M}}
\newcommand{\Ms}{M_{\odot}}
\newcommand{\m}{\langle}
\newcommand{\M}{\rangle}
\newcommand{\hf}{\tilde h}
\newcommand{\rf}{\tilde r}
\newcommand{\yf}{\tilde y}
\newcommand{\qf}{\tilde q}
\newcommand{\Qf}{\tilde Q}
\newcommand{\bml}{\begin{mathletters}}
\newcommand{\eml}{\end{mathletters}}
\def \Om {h_{100}^2\,\Omega}
\def \Opmin {h_{100}^2\,\Omega_p^{({\rm min})}}
\def \Omin  {h_{100}^2\,\Omega^{({\rm min})}}
\def \hrms {h_{\rm rms}}
\def \hc {h_{\rm c}}
\def\ltsima{$\; \buildrel < \over \sim \;$}
\def\simlt{\lower.5ex\hbox{\ltsima}}
\def\gtsima{$\; \buildrel > \over \sim \;$}
\def\simgt{\lower.5ex\hbox{\gtsima}}

\preprint{\vbox{\baselineskip=12pt
\rightline{RCG 00/2}}}
\title{High energy physics and the very early Universe with LISA}
\author{Carlo Ungarelli$^{(1,2)}$\thanks{Electronic address: carlo.ungarelli@port.ac.uk} 
and Alberto Vecchio$^{(2)}$
\thanks{Electronic address: vecchio@aei-potsdam.mpg.de}}
\address{${}^{(1)}$ School of Computer
Science and Mathematics, University of Portsmouth, Mercantile House,
Hampshire Terrace, Portsmouth P01 2EG, UK \\
${}^{(2)}$ Max Planck Institut f\"{u}r Gravitationsphysik,
Albert Einstein Institut\\
Am M\"{u}hlenberg 1, D-14476 Golm, Germany}


\maketitle

\begin{abstract}

Gravitational wave experiments will play a key role in the investigation
of the frontiers of cosmology  and the structure of fundamental fields
at high energies, by either setting stringent upper limits on, or
by detecting 
the primordial gravitational wave background produced in the early-Universe. 
Here we discuss the impact of space-borne laser interferometric detectors 
operating in the low-frequency window $\sim 10^{-6}$ Hz -- 1 Hz; the aim of
our analysis is to investigate whether a primordial background
characterized by a fractional energy density $\Om \sim 10^{-16}$ -- $10^{-15}$, which
is consistent with the prediction of "slow-roll" inflationary models, might
be detectable by the Laser Interferometer Space Antenna (LISA) 
or follow-up missions.

In searching for stochastic backgrounds, the presently planned LISA mission suffers 
from the lack of two detectors with uncorrelated noise. 
We analyze the sensitivity improvements that could be achieved
by cross-correlating the data streams from a pair of detectors of the LISA class;
we show that this configuration is extremely compelling, leading to
the detection of a stochastic background as weak as $\Om \simeq 5\times 10^{-14}$.
However, such instrumental sensitivity can not be fully exploited to measure
the primordial component of the background,
due to the overwhelming power of the signal produced by large populations
of short-period solar-mass binary systems of compact objects. 
We estimate that the primordial background can be observed only if its
fractional energy density $h_{100}^2\,\Omega$ is greater than $\approx 5\times 10^{-13}$.

The key conclusion of our analysis is that the stochastic radiation from unresolved
binary systems sets a fundamental limit on the sensitivity that can be achieved in
searching for the primordial background in frequencies between 
$\sim 10^{-6}$ Hz and 0.1 Hz, regardless of the instrumental noise level 
and the integration time. Indeed, 
the mHz frequency band, where LISA achieves optimal sensitivity, is
not suitable to probe slow-roll inflationary models.
We briefly discuss possible follow-up missions aimed at the frequency region
$\sim 0.1$ Hz -- 1 Hz, which is likely to be free from stochastic backgrounds
of astrophysical origin: no fundamental limits
seem to prevent us from reaching $\Om \sim 10^{-16}$, although the technological
challenges are considerable and deserve  careful study.
\end{abstract}

\bigskip

\narrowtext\noindent%

\draft

\section{Introduction}
\label{sec:int}

The Universe became "thin" to gravitational waves (GWs)
at the Plank epoch, corresponding to the cosmic time $\sim 10^{-43}$ sec.;
the gravitons decoupled from
the surrounding plasma at a temperature of the order of the Planck mass
$\sim 10^{19}$ GeV, and gravitational radiation produced at that epoch or later --
including the  electro-weak and the Grand Unification (GUT) scale -- 
has travelled undisturbed to us, carrying full 
information about the state of the Universe, and the physical processes from which
it took  origin. Indeed GW experiments will open radically 
new frontiers for cosmology and high energy  physics 
(see~\cite{Maggiore99,Creighton99} and reference therein for an extensive
discussion).

In the time frame $\sim$ 2002-2010 a large portion of the GW spectrum will progressively 
become accessible, mainly through large-scale laser interferometers. 
On the ground, the world-wide network of km-size interferometers
-- LIGO, GEO600, VIRGO and TAMA -- sensitive in the frequency band 
$\sim 10\,{\rm Hz} - 1\,{\rm kHz}$, will start carrying out "science runs" at the beginning
of 2002, with the realistic goal of directly detecting GW's. Several instrumental upgrades,
starting around 2005, will drive the sensitivity of the instruments to 
a GW stochastic background from $\Om \sim 10^{-6}$
(for the so-called initial generation) to $\Om \sim 10^{-10}$ 
(for the so-called advanced configuration). In space, 
a collaboration between ESA and NASA is carrying out the project called 
LISA (Laser Interferometer Space Antenna). This is a 
space-borne laser interferometer with arms of length
$5\times 10^6$ km, planned to fly by 2010~\cite{LISA_ppa}. 
This instrument guarantees the detection of GW's at low
frequencies ($\sim 10^{-5}\,{\rm Hz} - 10^{-2}\,{\rm Hz}$).

The purpose of this paper is to show the central role of the
experiments in the low-frequency window $\sim 10^{-6}$ Hz -- 1 Hz, 
with emphasis on instruments of the LISA class, in the search for the 
primordial GW background. Our aim is to identify the fundamental issues regarding
the achievement of a sensitivity in the range 
$h_{100}^2\,\Omega \sim 10^{-16}$ -- $10^{-15}$,
which is set by the theoretical prediction of "slow-roll" inflationary
models.

\subsection{The stochastic background spectrum}
\label{subs:spect}

A stochastic GW background is a random process that can be described only in terms
of its statistical properties. Without loss of generality, for the issues discussed
in this paper, we assume it to be isotropic, stationary, Gaussian
and unpolarized. The energy and spectral content of a stochastic background
are described by the dimensionless function
\be
\Omega (f) \equiv \frac{1}{\rho_{\rm c}}\,\frac{d\rho_{\rm gw}(f)}{d\ln f}\,;
\label{omegagw}
\ee 
$\rho_{\rm gw}$ is the energy density carried
by the background radiation, and
\ba
\rho_{\rm c} = \frac{3\,H_0^2\,c^2}{8\pi\,G_N} & \approx & 1.6\times 10^{-8}\,h_{100}^2
\,{\rm erg}/{\rm cm}^3\,,
\nonumber\\
& \approx & 1.2\times 10^{-36} \,h_{100}^2\,{\rm sec}^{-2}
\label{rhoc}
\ea
is the {\it critical energy density} required today to
close the Universe. The value of the Hubble constant (today) is 
\ba
H_0 & = & 100\,h_{100}\,\,{\rm km}\,{\rm sec}^{-1}\,{\rm Mpc}^{-1}\,,\nonumber\\
& \simeq & 3.2\times 10^{-18}\,h_{100}\,\,{\rm sec}^{-1}\,,
\label{H0}
\ea
where $h_{100}$ is known from observations to be in the range $0.4 \le h_{100}
\le 0.85$. 
$\Omega(f)$ is therefore the ratio of the GW energy density to the critical 
energy density per unit logarithmic frequency interval; one usually refers to 
$h_{100}^2\Omega(f)$, which is independent of the {\it unknown} value
of the Hubble constant.

It is useful to introduce the {\it characteristic amplitude} $\hc(f)$ of
the GW background: it is the
dimensionless characteristic value of the total GW background-induced fluctuation $h(t)$ at the output of an interferometer 
per unit logarithmic frequency interval:
\be
\m h^2(t) \M = 2 \int_0^\infty\, d(\ln f) \hc^2(f)\,;
\label{hc}
\ee
here $\m \,\M$ denotes the expectation value.
The spectral density $S(f)$ of the background is related to 
$\hc(f)$ by~\cite{Maggiore99}
\be
\hc^2(f) = 2 f S(f)\,,
\label{Sbg}
\ee
and $\Omega (f)$, $\hc(f)$, and $S(f)$ satisfy the relation~\cite{Maggiore99}
\be
\Omega(f) = \frac{2\pi^2}{3 H_0^2}\,f^2\,\hc^2(f) = 
\frac{4\pi^2}{3 H_0^2}\,f^3\,S(f)\,.
\label{OhS}
\ee
The characteristic amplitude over a frequency band $\Delta f$ is therefore:
\ba
\hc(f,\Delta f) & = & \hc(f)\,\left(\frac{\Delta f}{f}\right)^{1/2}\nonumber\\
& \simeq & 7.1\times 10^{-22}\,
\left[\frac{\Om(f)}{10^{-8}}\right]^{1/2}\,
\left(\frac{f}{1\,{\rm mHz}}\right)^{-3/2}\,
\left(\frac{\Delta_{\rm b} f}{3.2\times 10^{-8}\,{\rm Hz}}\right)^{1/2}\,,
\label{hc1}
\ea
where $\Delta_{\rm b} f \simeq 3.2\times 10^{-8}\,(1\,{\rm yr}/T)$ Hz is the width of the 
frequency bin for an observation time $T$. For comparison, the relevant characteristic 
amplitude of the LISA noise is $\sim 10^{-24}$.

\subsection{Sources of stochastic backgrounds}
\label{subs:sources}

The stochastic GW background can be divided into two broad classes, 
based on its origin: (i) the {\it primordial GW background} (PGB),
produced by physical processes in the early Universe, and (ii) the 
{\it astrophysically-generated GW background} (GGB), generated by the incoherent
superposition of gravitational radiation produced, at much later cosmic times, 
by a large number of astrophysical sources that can not be resolved individually. 
The emphasis of this paper is on the detectability of the PGB.

In this paper we will use the following conventions: $\Omega_p(f)$ and 
$\Omega_g(f)$ identify the fractional energy density in GW's, Eq.~(\ref{omegagw}),
carried by the primordial and the generated component of the GW background, respectively. If
no index is used, we refer to a general GW stochastic signal, with no assumption
about its production mechanism.

At present, there are three observational constraints on the PGB contribution to 
$\Omega (f)$: 
\begin{enumerate}
\item The high degree of isotropy of the cosmic microwave background
radiation sets a limit at ultra-low frequencies~\cite{Bennett96}:
\be
h_{100}^2\Omega_p(f) < 7\times 10^{-11}\,\left(\frac{f}{H_0}\right)^{-2}
\quad\,,\quad 
3\times 10^{-18}\,h_{100}\,{\rm Hz} \simlt f \simlt 10^{-16}\,h_{100}\,{\rm Hz}\,;
\ee
\item The very accurate timing of milli-second radio-pulsars 
constrains $\Omega_p(f)$ in a frequency range of the order of the
inverse of the observation time, typically of order
of a few years~\cite{KTR94}:
\be
h_{100}^2\Omega_p(f) < 10^{-8}
\quad\quad  f\sim 10^{-8}\,{\rm Hz}\,;
\ee
\item The standard model of big-bang nucleosynthesis constrains
the total energy content in GWs over a wide frequency
range~\cite{KT90}:
\be
\int_{f = 10^{-8}\,{\rm Hz}}^{\infty}\,h_{100}^2\Omega_p(f)\,d(\ln f) < 6\times 10^{-6}\,.
\ee
\end{enumerate}

To foresee what physical processes could have produced a detectable GW background
is an almost impossible challenge; nonetheless, it is enlighting to discuss some general
principles and possible generation mechanisms to show the typical sensitivity that experiments should achieve
in order to test different models. 

The main mechanisms that produce a PGB
can be divided into two broad categories (for a recent detailed review
see~\cite{Maggiore99}): (i) Parametric 
amplifications  of metric tensor perturbations that occurred 
during an inflationary epoch, and (ii) Some causal processes -- such as
phase transitions -- that took place in the early Universe.

Stochastic backgrounds produced by the parametric 
amplification  of metric tensor perturbations that occurs
during an inflationary epoch~\cite{Gri93} extend over a huge range of frequencies,
from $\sim 3\times 10^{-18}$ Hz up to a cutoff frequency in the GHz
range. In the window $\sim 10^{-16}\,\mbox{Hz} - 1\,\mbox{GHz}$, 
slow-roll inflationary models predict  a quasi scale-invariant 
spectrum whose typical magnitude -- in order to
satisfy the COBE bound -- cannot exceed $h_{100}^2\,\Omega_p\sim
10^{-14}$ in the LISA frequency band, as well as in the Earth-based detectors 
observational window~\cite{KW92}; a more refined  analysis~\cite{SRI}  
yields a more conservative upper limit: $h^2_{100}\,\Omega_p\sim 10^{-16} - 10^{-15}$. 
Superstring-inspired cosmological models~\cite{Veneziano91,GV93a,GV93b} predict a spectrum that,
for suitable choices of the free parameters of the model, could reach 
$h_{100}^2\Omega_p\sim 10^{-7}$ at the frequencies accessible either to 
Earth-based or to space-borne experiments, while satisfying the existing 
observational bound~\cite{BGGV95,BGV97,BMU97,AB97,UV00}. 

Stochastic backgrounds can also be produced by some 
classical causal processes that took place in the early Universe; for
this class of signals, the  characteristic frequency 
is related both to the time of emission and the corresponding
temperature ${\cal T}$. 
 
Non-equilibrium processes that occur at the reheating that takes place 
after inflation could provide a stochastic background
with cutoff frequency in the range $\sim 0.1\,{\rm mHz} - 1\,{\rm kHz}$,
corresponding to reheating temperatures between $\sim 1\,{\mbox TeV}$ and
$\sim 10^{9}\,{\mbox GeV}$. As an example, 
in hybrid and extended inflationary models, the exit
towards a radiation-dominated era is characterized by a first-order
phase transition, which, if strongly of the first order, generates a stochastic 
background with $h_{100}^2\,\Omega_p\sim 10^{-6}$
at frequencies that can vary from the LISA observational window up to the
sensitivity band of Earth-based interferometers ~\cite{Hyb}.

Phase transitions that inevitably occur at ${\cal T}\sim 10^2\,{\mbox MeV}$ 
(the QCD phase transition) and  ${\cal T}\sim 10^2\,{\mbox GeV}$ 
(the electroweak phase transition) produce GWs. In particular, if the electroweak
phase transition is  strongly of the first order, the spectrum 
is approximately $h_{100}^2\,\Omega_p\sim 10^{-11} - 10^{-9}$ at 
$f \sim 1\,{\rm mHz}$~\cite{KKT94};
the requirement of a strong first order phase transition, which is necessary in order to have
baryogenesis at the electroweak scale (see~\cite{RT99} and references
therein for a recent review), is directly related, in a minimal supersymmetric
extension of the standard model, to the mass of the super-partner of the top 
quark~\cite{CQW96,CQW98,LR98}. 
 
Cosmic strings, which are topological defects formed during phase
transitions, produce GW's whose typical frequency ranges from $f\sim
10^{-8}$ Hz up to $f\sim 10^{10}$ Hz with $h_{100}^2\,\Omega_p\sim 10^{-9} - 10^{-8}$,
see~\cite{BCS97} and references therein for a review.
 
Global phase transitions associated with 
some scalar field which acquires a non-zero vacuum expectation value (VEV) below a critical
temperature would produce, via a quite general 
relaxation process, GW's whose energy content is very significant,
$h_{100}^2\,\Omega_p\sim 10^{-6}$, for VEV's near the Planck/string 
scale~\cite{Krauss92,Hogan98}. 

Recently there has been a great amount of
theoretical activity investigating higher dimensional
"brane-world" scenarios, where gravity begins to probe the extra
dimensions at energies as low as $10^3$ GeV; and an estimate of 
possible GW backgrounds in such models was presented recently 
in~\cite{Hogan00}.

These examples clearly show that the investigation of 
the primordial GW stochastic background in the low-frequency regime would provide 
us key information about the physics beyond the standard model and/or could allow us
to discriminate between different inflationary cosmological
models.

\subsection{Detecting a stochastic background}
\label{subs:det}

A stochastic background is a random process which is intrinsically indistinguishable
from the detector noise. In order to 
detect such a signal, the optimal signal processing strategy calls for correlations 
between two (or more pairs of) instruments, possibly widely separated in order to minimize
the effects of {\it common} noise sources. The relevant
data analysis issues 
have been thoroughly addressed in~\cite{Flanagan93,AR99}; here
we simply review the main concepts, and refer to~\cite{Flanagan93,AR99},
and references therein, for more details. 

The statistical analysis is based on the following assumptions:
the signal and the detector noise
are uncorrelated; the noise in each detector is stationary and Gaussian, and 
possible noise correlations between two detectors are negligible.

We define the output (signal + noise) of the two instruments as
$o_1(t)$ and $o_2(t)$; the cross-correlation signal $C$ that one constructs is therefore
of the form:
\be
C \equiv \int_{-T/2}^{T/2}\,dt \int_{-T/2}^{T/2}\,dt'\, o_1(t)\, o_2(t')\,Q(t-t')\,,
\label{corr}
\ee
where $Q(t-t')$ is a suitable filter function. In the general case, 
the filter function depends on $t$ {\it and} $t'$
independently, that is $Q = Q(t,t')$; here we have used the property of the
signal of being stationary, and therefore $Q(t,t') = Q(t-t')$. The SNR is 
defined as:
\be
{\rm SNR} = \frac{\mu}{\sigma}\,,
\label{snr}
\ee
where $\mu$ and $\sigma$ are the mean value and the variance of the observable  $C$:
\be
\mu \equiv \m C\M = T\,\left(\frac{3 H_0^2}{20\pi^2}\right)\,\left(\tilde Q,\tilde A\right)\,,
\label{mu}
\ee
\be
\sigma^2 \equiv \m C^2\M -  \m C\M^2 = \frac{T}{4}\,\left(\tilde Q,\tilde Q\right)\,.
\label{sigma}
\ee
Eqs.~(\ref{mu}) and~(\ref{sigma}) are written in terms of the usual 
{\it inner product}~\cite{AR99}
\be
(a,b) \equiv \int_{\rm -\infty}^{+\infty} \,df\,\tilde a^*(f) \tilde b(f) R(f)\,,
\label{ip}
\ee
where $\tilde Q(f)$ is the Fourier transform of $Q(t-t')$. 
The functions $R(f)$ and $\tilde A(f)$ are defined as follows:
\ba
R(f) & \equiv & S_n^{(1)}(f) S_n^{(2)}(f)\times \nonumber \\
&&\left\{
1 + \left(\frac{3 H_0^2}{10\pi^2}\right)\,
\frac{\Omega(f)}{f^3}\,\left[\frac{S_n^{(1)}(f) + S_n^{(2)}(f)}{S_n^{(1)}(f) S_n^{(2)}(f)}
\right]+
\left(\frac{3 H_0^2}{10\pi^2}\right)^2
\frac{\Omega^2(f)\,\left[1 + \gamma(f)^2\right]}{f^6\,S_n^{(1)}(f)S_n^{(2)}(f)}\right\}\,,
\label{Rf}
\ea
\be
\tilde A(f) \equiv \frac{\gamma(f) \Omega(f)}{f^3 R(f)}\,.
\label{Af}
\ee
In Eq.~(\ref{Rf}), $S_n^{(k)}(f)\,,k=1,2$ is the one-sided noise power 
spectral density of the $k-$th detector, and $\gamma(f)$ is the so-called 
{\it overlap reduction function}, which depends entirely on the relative
orientation and location of the two detectors; it accounts for 
SNR losses that occur when the instruments are not optimally located and 
oriented, cf. Eq.~(\ref{snr_w}) and Sec.~\ref{sec:overlap}.

Using Eqs.~(\ref{mu}) and~(\ref{sigma}), one can cast Eq.~(\ref{snr}) in the form:
\be
{\rm SNR}^2 = T\,\left(\frac{3 H_0^2}{10\pi^2}\right)^2\,
\frac{\left(\tilde Q,\tilde A\right)^2}{\left(\tilde Q,\tilde Q\right)}\,.
\label{snr1}
\ee
The optimal choice of the filter $\tilde Q$, is thus based on the 
maximizing the SNR, Eq.~(\ref{snr1}), and is given by:
\be
\tilde Q(f) = ({\rm const.}) \times \tilde A(f)\,,
\label{Qf}
\ee
where the overall normalization factor is arbitrary. 
Note that Eqs.~(\ref{corr})-(\ref{Qf}) are valid for a background of arbitrary
energy density $\Omega(f)$. In the case of a signal much weaker than the noise, 
$H_0^2\Omega(f)/f^3 \ll S_n^{(k)}(f)$, one can Taylor expand Eqs.~(\ref{Rf})
and~(\ref{Af}), retaining only the leading order term. As a consequence,
Eq.~(\ref{snr1}) reduces to:
\be
{\rm SNR} \simeq \frac{3 H_0^2}{\sqrt{50}\pi^2}\,T^{1/2}\,
\left[\int_{0}^{\infty}\,df\,
\frac{\gamma(f)^2\Omega^2(f)}{f^6\,S_n^{(1)}(f)S_n^{(2)}(f)}\right]^{1/2}
\quad\quad ({\rm signal} \ll {\rm noise} )\,.
\label{snr_w}
\ee
It is convenient to introduce the noise characteristic amplitude $\hrms$,
equivalent to $\hc$, as follows: 
\ba
\m n^2(t)\M & = & \int_0^\infty\, df S_n(f) \nonumber\\
& = & 2 \int_0^\infty\, d(\ln f) \hrms^2(f) \,.
\label{nav}
\ea
It is enlightening to write Eq.~(\ref{snr_w}), using Eqs.~(\ref{hc}) and~(\ref{nav}),
in the form:
\be
{\rm SNR} \sim \gamma(f_c) \left(\Delta f\,T\right)^{1/2}\,
\left[\frac{\hc(f_c)}{\hrms(f_c)}\right]^2\,;
\label{snr_appr}
\ee
here we have assumed that the frequency band $\Delta f$, which contains
most of the SNR, is centered on the 
characteristic frequency $f_c$, and is sufficiently
small; the noise spectral density of the two instruments, that for
simplicity we assume identical, and the overlap reduction function can be therefore
treated as 
roughly constant. If only one instrument is in operation, one could in principle
detect a stochastic background with SNR$\simgt 1$ when $\hc \simgt \hrms$;
with two instruments one can detect the signal when 
$\hc \simgt \hrms/[\gamma(f_c)\,(\Delta f\,T)^{1/4}]$.
Cross-correlation experiments are therefore highly desirable for
both detection confidence and sensitivity. In fact, one can isolate the stochastic 
GW signal from all the spurious contributions which 
are uncorrelated between the two instruments. Common noise sources, which 
correlate on the same light-travel time scale, might, however, be  present,
degrading the overall sensitivity. Moreover, GW signals are expected to be very 
weak, well buried into the noise; using cross-correlations, through optimal
filtering one increases the sensitivity by a factor 
$\sim 10$ $(\Delta f/1\,{\rm mHz})^{1/4}$ $(T/10^7\,{\rm sec})^{1/4}$, 
with respect to the single detector case. 

\subsection{Summary of the results}
\label{subs:res}

The goal of this paper is to explore the capability of space-borne
laser interferometers, such as LISA and its successors,
in searching for the primordial GW stochastic backgrounds.
The analysis of the LISA technology leads us to the following conclusions:

\begin{itemize}

\item A PGB of fractional energy density $h_{100}^2\Omega_p \simgt 10^{-10}$ 
definitely shows up as an excess power component in the data of a single LISA 
interferometer over a large frequency window; it might be detectable by
calibrating the noise-only response of the instrument, but the issue
of decisively assigning this contribution to a real primordial signal 
remains open.

\item Cross-correlations between the data streams of two identical LISAs, 
characterized by the presently
estimated instrumental noise, allow us to reach a minimum value of the fractional
energy density carried by GWs in the range
$5\times 10^{-14} \simlt \Omin\simlt 10^{-12}$ for an integration 
time $T = 10^7$ sec, 
depending on the location and orientation of the two detectors. 

\item Such remarkable sensitivity, however, does not apply 
to {\it primordial} GW backgrounds; in
fact, the copious number of short period solar-mass binary systems in the Universe
produce a GGB that overwhelms the PGB in the key mHz region; 
we estimate that {\it the minimum
detectable value of the primordial GW background 
is $\Opmin \simgt 5\times 10^{-13}$}.

\item The cross-correlation of the data streams from two LISA detectors 
provides, therefore,  a powerful tool to extract information about populations 
of binary systems of short-period solar-mass compact objects in the Universe:
the GGB is detectable at signal-to-noise ratio $\sim 100$.

\end{itemize}

We would like to emphasize that the third generation Earth-based laser interferometers,
instruments such as advanced LIGO (LIGO III) and 
EUGO, a new European detector currently under study,
will be able to achieve a sensitivity 
$\Omin \sim 10^{-10}$; space-based interferometers will therefore
play a primary role in GW cosmology.

Although the former results are very encouraging, our analysis
leads to the rather obvious, but somewhat disappointing outcome that experiments
in the frequency band
$\sim 10^{-6}\, {\rm Hz} - 0.1\,{\rm Hz}$ will be limited to a sensitivity of the order 
$\Om_p \sim 10^{-13} - 10^{-12}$; this limit can not be improved 
by reducing the instrumental noise and/or increasing the integration time;
in fact, GGB's produce a residual correlation in the filter output designed to detect
the PGB, that can not be eliminated: {\it the
mHz frequency window is not suitable to search for a primordial
gravitational wave background characterized by $\Om_p < 5\times 10^{-13}$}.

It is therefore worth asking whether future technological developments, and
more ambitious missions will enable the detection of a very weak PGB.
Our present understanding suggests that cross-correlation experiments carried 
out with a pair of space-based interferometers with optimal sensitivity in the band 
$\sim 0.1\,{\rm Hz} - 1\,{\rm Hz}$ could be able to meet the target 
$\Om_p \sim 10^{-16}$. In this frequency band, in fact, astrophysically
generated backgrounds are not present, and an experiment is limited in principle
only by the instrumental noise. The design of a detector aimed at the window
$\sim 0.1\,{\rm Hz} - 1\,{\rm Hz}$ imposes stringent requirements 
on the power and frequency of the laser, as well as on the dimensions of the "optics" 
and on several other components of the instrument. In order to test slow-roll inflation 
an effective -- {\i.e.} after residual radiation from individual binary 
systems has been removed -- rms noise level $\sim 10^{-24}$, 
and rather long integration times ($\approx 3$ yrs) are required. Such technological
and data analysis issues have been little investigated so far, and deserve
careful consideration.

%
%

The paper is organized as follows. In Sec.~\ref{sec:overlap} we derive a 
closed form expression of the overlap reduction function for a pair of 
space-based interferometers. In Sec.~\ref{sec:gb} we review our present astrophysical
understanding of the GW background generated by the incoherent superposition
of radiation emitted by galactic and extra-galactic short period solar-mass binary
systems. In Sec.~\ref{sec:sens} we present the key results of the paper:
we estimate the sensitivity that can be achieved by the present LISA
technology, and possible follow-up missions, by cross-correlating
the outputs of two identical detectors with uncorrelated noise.
Sec.~\ref{sec:concl} contains our conclusions and pointers to future work.

\section{The overlap reduction function for LISA detectors}
\label{sec:overlap}

In this section we compute a closed form expression for the 
overlap reduction function $\gamma(f)$, see Eqs.~(\ref{Rf})-(\ref{Qf}), 
for space-borne interferometers, and in particular for instruments
characterized by a LISA-like orbital configuration. Indeed, our analysis provides
explicit formulae that can be directly applied, with little changes, to any
orbital configuration.

The overlap reduction function is a dimensionless function of the 
frequency $f$, which measures the degradation of the SNR when the detectors
are not optimally oriented and located. At any given frequency, $\gamma(f)$
depends entirely on the relative separation and orientation of the 
instruments, and its behavior has a very intuitive physical explanation.
It is maximum when the detectors are co-located and co-aligned.
As the detectors (parallel to each others) are shifted apart, at a distance
$D$, the signal drives oscillations in the two instruments that are progressively 
out of phase; there is an effective correlation only if the separation is smaller
than approximately half of the characteristic
wavelength of the gravitational radiation; equivalently, the frequency
band over which one accumulates SNR is
\be
f \simlt 1\,\left(\frac{D}{1\,{\rm AU}}\right)^{-1} \,{\rm mHz}\,,
\ee
where one Astronomical Unit (AU) corresponds to $1.4959787\times 10^{13}$ cm.
For two coincident detectors, $\gamma(f)$ decreases, at any frequency, as one 
instrument is rotated with respect to the other, because the detectors are
excited in different ways by the different polarizations;
for a rotation of $\pi/4$, the overlap reduction function is identically zero. 
In the general case, $\gamma(f)$ shows a complex behavior which is the 
superposition of the two effects that we have just described.

\subsection{The LISA mission}
\label{subs:lisa}

LISA is an all-sky monitor with a quadrupolar antenna pattern. Its orbital
configuration was conceived in order to keep the geometry of the interferometer
as stable as possible during the mission, as well as to provide an optimal coverage
of the sky: a constellation 
of three drag-free spacecraft (containing the free-falling test masses) 
is placed at the vertices of an ideal equilateral triangle 
with sides $\simeq 5\times 10^6\,{\rm km}$; it forms
a three-arm interferometer, with a $60^\circ$ angle between two 
adjacent laser beams. The LISA orbital motion is such that
the barycentre of the instrument is inserted in a heliocentric (essentially circular)
orbit, following by $20^\circ$ the Earth; the detector plane is tilted by 
$60^\circ$ with respect to
the Ecliptic and the instrument counter-rotates around the normal to the 
detector plane with the same period of $1\,{\rm yr}$. 

We introduce a Cartesian reference frame
$(x,y,z)$ tied to the Ecliptic, with ${\bf \hat z}$ perpendicular to 
the Ecliptic plane, and ${\bf \hat x}$ and ${\bf \hat y}$ in the plane 
itself, and oriented in such a way to define a left-hand tern. In this frame, 
LISA's centre-of-mass is described by the polar angles
\be
\Theta = \frac{\pi}{2}\quad , \quad \Phi(t) = \Phi_0 + n_{\oplus} t
\quad,\quad\quad \left(n_{\oplus}\equiv \frac{2\pi}{1\,{\rm yr}}\right)\,;
\label{barlisa}
\ee
$\Phi_0$ sets the position of the detector barycentre at some arbitrary reference time. 
The time evolution of the unit vectors ${\bf \hat l}_j$ ($j = 1,2,3$) 
along each arm are described by the following expression~\cite{Cutler98}:
\ba
{\bf \hat l}_j & = &
\left[\frac{1}{2}\,\sin\alpha_j(t)\,\cos\Phi(t) - \cos\alpha_j(t)\,\sin\Phi(t)\right]
\,{\bf \hat x} \nonumber\\
&& +
\left[\frac{1}{2}\,\sin\alpha_j(t)\,\sin\Phi(t) + \cos\alpha_j(t)\,\cos\Phi(t)\right]
\,{\bf \hat y} 
+ \left[\frac{\sqrt{3}}{2}\,\sin\alpha_j(t)\right]\,{\bf \hat z} \,,
\label{larm}
\ea
where $\alpha_j(t)$ increases linearly with time, according to
\be
\alpha_j(t) = n_{\oplus} t - (j-1)\pi/3 + \alpha_0 \;,  
\label{alphaj}
\ee
and $\alpha_0$ is just a constant specifying the orientation of
the arms at the arbitrary reference time $t=0$.

In the next section we will derive the expression of the overlap reduction
function for two detectors characterized by a LISA-like motion. 
The time dependence of ${\bf \hat l}_j$ and the
centre-of-mass are described by Eqs.~(\ref{barlisa}),
(\ref{larm}) and~(\ref{alphaj}) for both instruments, just with different 
initial conditions $\alpha_0$ and $\Phi_0$. We will use the notation $\alpha_{01}$
and $\Phi_{01}$, and $\alpha_{02}$ and $\Phi_{02}$ to indicate the corresponding values
of the detector "1" and "2", 
respectively. For future convenience, we also introduce the notation:
\be
\Delta \Phi_0 = \Phi_{02}-\Phi_{01} \quad,\quad
\Delta \alpha_0 = \alpha_{02}-\alpha_{01}\,.
\label{deltaap}
\ee%
\narrowtext\noindent%
\begin{figure}[htb!]
\begin{center}
{\epsfig{file=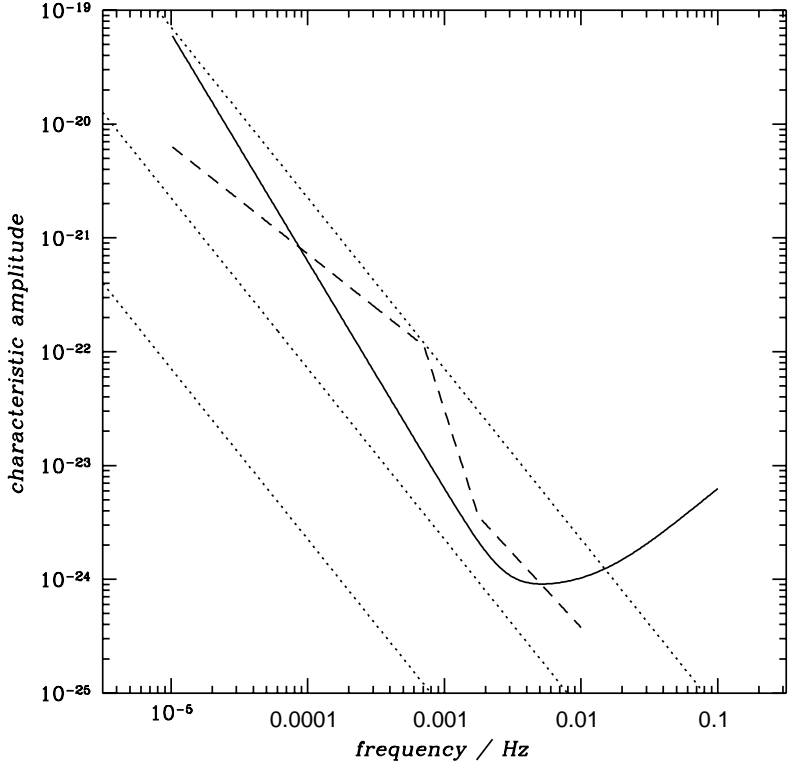,
angle=0,width=5.0in,bbllx=25pt,bblly=50pt,bburx=590pt,bbury=740pt
}}
\caption{\label{fig:noise} 
The characteristic amplitude of stochastic gravitational wave backgrounds
of primordial and astrophysical origin and the LISA rms noise amplitude.
The plot shows the LISA rms noise amplitude in one year of observation (solid line),
as a function of frequency, and compares it to the characteristic amplitude of
several stochastic signals: a backgrounds with flat spectrum 
$h_{100}^2\,\Omega = 10^{-10}$, $10^{-13}$, $10^{-16}$ (dotted-lines, 
from top to bottom, respectively), and the astrophysically generated
galactic background (dashed-line), according to the estimate given by
Bender and collaborators, cfr. Eq.~({\protect\ref{Sncg}}).
%
}
\end{center}
\end{figure}
\narrowtext\noindent%

The main noise sources that affect the mission have been addressed by the LISA 
Science Team and yield the following expression for the
expected noise spectral density~\cite{LISA_ppa}
\be
S_n(f) \simeq 10^{-48}\,\left[
1.22\times 10^{-3}\, f^{-4} + 16.74 + 9.7\times 10^4\, f^2\right]\,{\rm Hz}^{-1}\,;
\label{Sni}
\ee
the rms noise amplitude that is inferred from $S_n(f)$ is shown in 
Fig.~\ref{fig:noise}.

\subsection{The overlap reduction function}
\label{subs:gammaL}

To derive a closed form expression of the overlap reduction 
function $\gamma(f)$, we follow the formalism developed by Allen and Romano~\cite{AR99},
which in turn was based on the analysis done by Flanagan~\cite{Flanagan93}.
$\gamma(f)$ is formally given by
\be
\gamma(f) = \rho_1(x)\,d^{(1)}_{ab}d^{(2)ab}+
\rho_2(x)\,d^{(1)}_{ab}S^b d^{(2)ac}S_c +
\rho_3(x)\,d^{(1)}_{ab}d^{(2)}_{cd}S^aS^bS^cS^d \,.
\label{gamma}
\ee
In the previous expression, $d^{(k)}_{ab}$ are the {\it detector response tensors},
defined as:
\be
d^{(k)}_{ab}=\frac{1}{2}\,\left[m_a^{(k)}m_b^{(k)}-n_a^{(k)}n_b^{(k)}\right]\,,
\label{d12}
\ee 
where $k=1,2$ labels the instrument, and  
${\bf \hat m}^{(k)}$ and ${\bf \hat n}^{(k)}$ are the unit vectors along the arms 
of the interferometers; our (arbitrary) choice corresponds to
${\bf \hat l}_1$ and ${\bf \hat l}_2$, respectively, given by 
Eqs.~(\ref{larm}) and~(\ref{alphaj}), with the appropriate initial conditions.
The unit vector along the direction that connects the centers of mass of the 
two detectors is
\be
{\bf \hat S} = 
\frac{\cos\Phi_2-\cos\Phi_1}{\sqrt{2(1-\cos(\Phi_2-\Phi_1))}}\, {\bf \hat x} 
+ \frac{\sin\Phi_2-\sin\Phi_1}{\sqrt{2(1-\cos(\Phi_2-\Phi_1))}}\,{\bf \hat y}\,;
\label{S}
\ee
$x$ is a dimensionless parameter defined as
\be
x\equiv \frac{2\pi f D}{c}\,,
\label{x}
\ee
where
\be
D = R_{\oplus}\sqrt{2\,\left(1 - \cos\Delta\Phi_0\right)}\,,
\label{D}
\ee
and the functions $\rho_j(x)$ are given by
\ba
\rho_1(x) & = & 5\,j_0(x) - \frac{10}{x}\,j_1(x) + \frac{5}{x^2}\,j_2(x)\,, \nonumber\\
\rho_2(x) & = & -10\,j_0(x) + \frac{40}{x}\,j_1(x) - \frac{50}{x^2}\,j_2(x)\,, \nonumber\\
\rho_3(x) & = & \frac{5}{2}\,j_0(x) - \frac{25}{x}\,j_1(x) + \frac{175}{2 x^2}\,j_2(x)\,, 
\label{rho}
\ea
where
\ba
j_0(x) & = & \frac{\sin x}{x}\,, \nonumber\\
j_1(x) & = & \frac{\sin x}{x^2} - \frac{\cos x}{x}\,, \nonumber\\
j_2(x) & = & 3\,\frac{\sin x}{x^3} - 3\,\frac{\cos x}{x^2} - \frac{\sin x}{x} \,,
\label{j}
\ea
are the standard spherical Bessel functions.%
\narrowtext\noindent%
\begin{figure}[htb!]
\begin{center}
{\epsfig{file=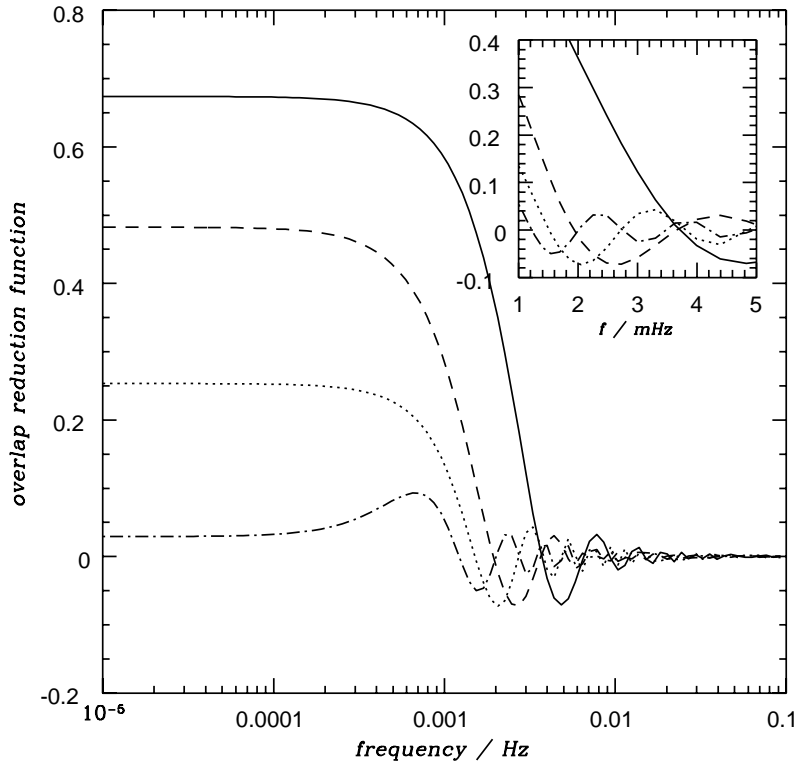,
angle=0,width=5.0in,bbllx=25pt,bblly=50pt,bburx=590pt,bbury=740pt
}}
\caption{\label{fig:gf} The overlap reduction function $\gamma(f)$ for
a pair of identical laser interferometers with the LISA orbital configuration.
The plot shows $\gamma(f)$ as a function of the frequency for four different
centre-of-mass separation angles $\Delta\Phi_0$, which correspond
to different distances of the two detectors, cfr. Eq.~({\protect{\ref{D}}}):
$\Delta\Phi_0 = 20^\circ$ (solid line), $\Delta\Phi_0 = 40^\circ$ (dashed line), 
$\Delta\Phi_0 = 60^\circ$ (dotted line), $\Delta\Phi_0 = 90^\circ$ (dotted-dashed line).
Here we always assume that the initial orientation of the detectors is such
that $\Delta\alpha_0=0$. The small panel at the top zooms the behavior of $\gamma(f)$
in the mHz region.
}
\end{center}
\end{figure}
\narrowtext\noindent%

By substituting Eq.~(\ref{larm}) into Eq.~(\ref{d12}), and
combining it with Eqs.~(\ref{S})-(\ref{j}), one can derive the following expressions:
\ba
d^{(1)}_{ab}d^{(2)ab} & = & \sum_{k=0}^{4}\, [A_k\,\cos(2\Delta\alpha_0+k\Delta\Phi_0)+
B_k\,\sin(2\Delta\alpha_0+k\Delta\Phi_0)\,, \\
\label{t1}
d^{(1)}_{ab}S^b d^{(2)ac}S_c & = &
\sum_{k=0}^{4}\,[C_k\,\cos(2\Delta\alpha_0+k\Delta\Phi_0)+
D_k\,\sin(2\Delta\alpha_0+k\Delta\Phi_0)]\,, \\
\label{t2}
d^{(1)}_{ab}d^{(2)}_{cd}S^aS^bS^cS^d & = &
\sum_{k=0}^{4}\,[E_k\,\cos(2\Delta\alpha_0+k\Delta\Phi_0)+
F_k\,\sin(2\Delta\alpha_0+k\Delta\Phi_0)]\,,
\label{t3}
\ea
where $\Delta\alpha_0$ and $\Delta\Phi_0$ are given by Eq.~(\ref{deltaap}), and 
$A_k$, $B_k$, $C_k$, $D_k$, $E_k$ and $F_k$ (for $k = 0,..,4$) are numerical
coefficients, that are given in Appendix A. Inserting Eqs.~(\ref{t1})-(\ref{t3}) 
in Eq.~(\ref{gamma}), the overlap reduction function becomes:
\be
\gamma(f) = \sum_{k=0}^{4}\,
\left[P_k(x) \,\cos\varphi_k(\Delta \alpha_0,\Delta \Phi) +
Q_k(x) \,\sin\varphi_k(\Delta \alpha_0,\Delta \Phi) \right]
\label{gamma1}
\ee
where
\ba
P_k(x) & = & \rho_1(x)\,A_k + \rho_2(x)\,C_k +  \rho_3(x)\,E_k\,,\nonumber\\
Q_k(x) & = & \rho_1(x)\,B_k + \rho_2(x)\,D_k +  \rho_3(x)\,F_k\,,
\label{PQ}
\ea
depend only on the detector separation and the radiation frequency, and
\be
\varphi_k(\Delta\alpha_0,\Delta\Phi_0) 
\equiv 2\,\Delta\alpha_0 + k\,\Delta\Phi_0\,,
\label{dphi}
\ee
is a function of the relative orientation of the instruments.%
\narrowtext\noindent%
\begin{figure}[htb!]
\begin{center}
{\epsfig{file=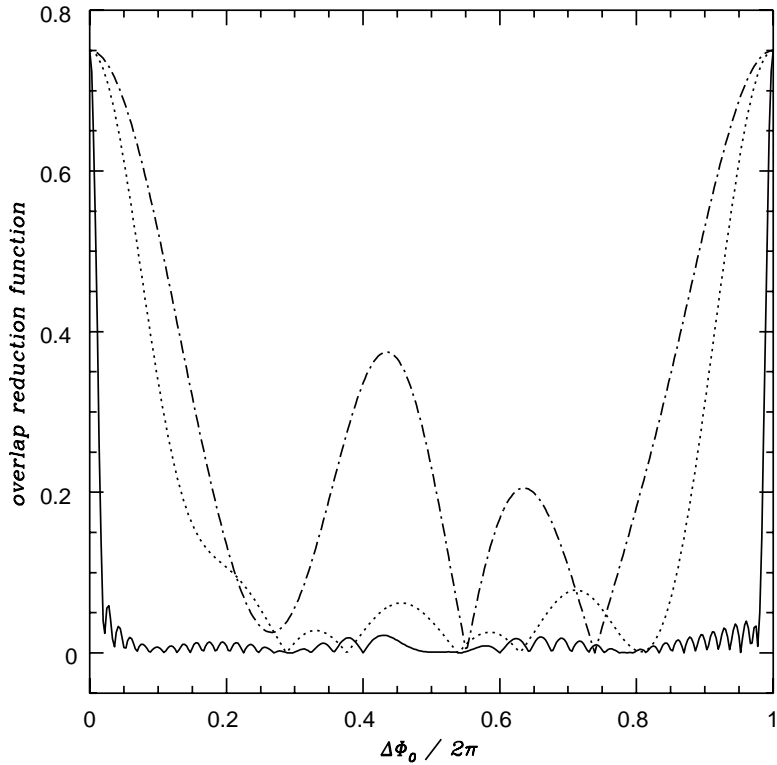,
angle=0,width=5.0in,bbllx=25pt,bblly=50pt,bburx=590pt,bbury=740pt
}}
\caption{\label{fig:gp} The overlap reduction function for
a pair of identical laser interferometers with the LISA orbital configuration.
The plot shows $\gamma(f)$ as a function of centre-of-mass 
separation angle $\Delta\Phi_0$ (in units $2\pi$), for selected frequencies:
$f = 10^{-2}\rm Hz$ (solid line), $f = 10^{-3}\rm Hz$ (dotted line), and
$f = 10^{-4}\rm Hz$ (dotted-dashed line). The constants
$\alpha_{01}$ and $\alpha_{02}$ are selected in such a way that, at the (arbitrary) 
reference time $t = 0$, $\Delta\alpha_0 = 0$. 
}
\end{center}
\end{figure}
\narrowtext\noindent%

We would like to stress that our definition is such that $\gamma(f) = 1$ $\forall f$,
{\it for co-aligned and co-located interferometers with arms perpendicular to each
others}, that is the angle between ${\bf \hat m}$ and ${\bf \hat n}$
is $\pi/2$. However, the LISA opening angle is $\pi/3$, and the detector response
is reduced by the factor $\sqrt{3}/2$: 
as a consequence, the maximum value that $\gamma(f)$
can attain is $3/4$. 

Notice also that, as pointed out by Cutler~\cite{Cutler98},
the read-outs from the three arms of LISA can be combined in such
a way to form the outputs, say $o_I$ and $o_{II}$, of two co-located interferometers 
rotated by $\pi/4$ one with respect to the other, 
whose noise is uncorrelated at all frequencies. Unfortunately, the cross-correlation
of $o_I$ with $o_{II}$ is useless for searching for stochastic backgrounds, as
the overlap reduction function is identically zero over the whole frequency range.

Figs.~\ref{fig:gf} and~\ref{fig:gp} show the behavior of $\gamma(f)$ as a function 
of the frequency and the separation angle $\Delta \Phi_0$, which is
equivalent to the distance $D$, see Eq.~(\ref{D}). Here we assume
that the instruments are inserted into their orbits so that $\Delta \alpha_0 = 0$. 
In this case, at fixed frequency, the overlap reduction function depends only on 
$\Delta \Phi_0$, as the relative orientation of the detectors is determined
only by $\Phi(t)$, cf. Eqs.~(\ref{larm}) and~(\ref{alphaj}). 
This also implies that as the two instruments are
moved apart -- $\Delta\Phi_0$ increases -- their orientation changes
too. In order to achieve the maximum $\gamma(f)$ for a given separation,
one should therefore suitably tune the initial orientation of the detector arms. 
The plots clearly show that at low frequencies, 
$f\simlt 5\times 10^{-4}$ Hz, $\gamma(f)$ is fairly flat and close
to its maximum value (which is set by the separation and orientation);
in fact the radiation wavelength is 
$\lambda_{\rm gw} \simeq 2 (f/1\,{\rm mHz})^{-1}$ AU,
and for separations smaller than 1 AU, the degradation of
SNR at very low-frequencies is less than a factor $\approx 2$. At high
frequencies, say $f\simgt 10^{-2}$ Hz, placing two interferometers at a
distance larger than $10^{6}$ km, would severely degrade the SNR,
by a factor of 10 or more.

\section{The astrophysically generated stochastic background}
\label{sec:gb}

The so-called {\it astrophysically generated GW stochastic background} 
(GGB) is mainly due to the incoherent
superposition of gravitational radiation emitted by short-period
solar-mass binary systems. A variety of binary populations contribute to it,
but the main contribution, in the mHz region, is due to close white-dwarf binaries
(CWDB's). Present estimates suggest that it is above the 
LISA instrumental noise in the frequency region 
$\approx 10^{-4} - 3\times 10^{-3}$ Hz,
right at the heart of the observation window, see Fig.~\ref{fig:noise}. 
However, sizeable effects are also given by other sources such as
W UMa (Ursae Majoris) binary stars, 
unevolved binaries, cataclysmic binaries, neutron star-neutron star (NS-NS) 
binary systems, black hole-neutron star (BH-NS) binaries, and possibly BH-BH 
binary systems.

The GGB is a {\it guaranteed} GW source in the low-frequency band; however, 
it is likely to overwhelm the PGB, degrading the sensitivity of the instruments 
in searching for a stochastic 
signal produced in the early Universe. A rigorous analysis 
of GGB's goes far beyond the purpose of this paper; here we review the main 
features and discuss the fundamental theoretical issues.
We refer the reader to~\cite{HBW90,BH97,PP98,KP98,SFMP00}, and references therein, 
for a thorough discussion of the astrophysical sources.

Ultimately, the reason that the radiation generated by large populations
of binary systems is effectively a stochastic signal is simple: there are 
too many free parameters that one needs to fit the data in order to {\it resolve}
all the binary systems that contribute to the signal. 
In fact, our galaxy contains $\sim 10^7$ CWDB's; they evolve,
due to radiation reaction, over a time scale $\simgt 10^7$
years. Therefore, during the typical observation time $T
\sim 1$ yr, they are seen as highly monochromatic, and, 
in the band $10^{-4} - 10^{-3}$ Hz,
each frequency bin is "contaminated" by roughly $10^3$ sources. 
The problem of resolving each individual binary system is actually made  worse 
by the motion of the detector, because,
in addition to the frequency, one needs to solve also for the source position
in the sky and the orientation of the orbital plane.

In the following we discuss the estimates $\Omega_g(f)$
of the spectral energy density of generated backgrounds, and in particular 
their {\it isotropic} component, and the critical frequency $f_g$ 
up to which GGB's are indeed present. For $f\simgt f_g$, 
the individual binary sources of astrophysical populations can be resolved, and 
their radiation subtracted from the data, opening up the band to search for a PGB.
The former two properties are essential in addressing the possibility of detecting
PGBs with space-based experiments.

In this context, it is useful to divide
the sources that contribute to the GGB into two categories: 
galactic sources and extra-galactic sources. Their key distinguishing feature
of relevance here is the degree of isotropy of the GGB that they generate, 
for an observer on board a LISA-like detector. In fact, the
extra-galactic contribution is expected to be isotropic
to a rather high degree (the radiation being dominated by binary systems 
at cosmological distances; for more details see~\cite{KP99}); it is, therefore,
impossible to discriminate it from the PGB. On the contrary, the GGB produced by 
galactic sources is clearly highly anisotropic. In fact, galactic stars are spatially 
distributed, approximately, according to $\exp(-r/r_0)\,\exp[-(z/z_0)^2]$, 
where $r$ is the radial distance to the Galactic centre, $r_0 \simeq 5$ kpc, 
and $z$ the height above the Galactic plane; $z_0\sim 5$ kpc for neutron star binaries
and less than 300 pc for the other types. 
Due to the peripheral location of the Solar System, and the change of orientation
of the LISA arms during the years-long observation time, 
galactic generated backgrounds appear strongly 
anisotropic~\cite{GP97}. It is also conceivable that the isotropic
portion of the galactic contribution does not exceed the total extra-galactic
GGB.

Several astrophysical uncertainties affect the estimates of the GGB that 
have been carried out so far. A careful analysis of the galactic 
contribution has been performed by~Bender and collaborators~\cite{HBW90,BH97},
taking into account a wide range of binary populations. A
good fit of the galactic GGB spectral density is:
\be
S_g(f) =
\left\{\begin{array}{ll}
10^{-42.685}\, f^{-1.9} & 10^{-5} \le f \le 10^{-3.15}\,,\nonumber\cr
10^{-60.325}\, f^{-7.5} & 10^{-3.15} \le f \le 10^{-2.75}\,,\nonumber \cr
10^{-46.85}\, f^{-2.6}  & 10^{-2.75}\,\le f\,, \\
\end{array}
\right.
\label{Sncg}
\ee
which is related to $\Omega_g(f)$ by Eq.~(\ref{OhS}).
In Eq.~(\ref{Sncg}) the space density of CWDBs is assumed to be $10\%$ of the
theoretical value predicted by Webbink~\cite{Webbink84}, and 
the radiation from helium cataclysmic variables, likely to contribute significantly
in the frequency window $\simeq 1\,{\rm mHz} - 3$ mHz, is not taken into account. 
The estimate~(\ref{Sncg}) can be definitely regarded as a solid lower limit, and 
is likely correct within to a factor $\sim 3$; see also~\cite{PP98,KP98}.
We would also like to stress that the dominant contribution from CWDBs switches 
off at $\sim 10^{-2}$ Hz, where the binary systems coalesce; 
for $f \simgt 10^{-2}$ Hz, NS-NS binary systems are the sources that contribute 
most to the GGB.

The extragalactic contribution to the GGB has been estimated in~\cite{KP98},
and is weaker than the galactic contribution by a factor $\approx$ 10-to-3 in the
relevant frequency band, where the main uncertainty comes from 
the star formation rate at high redshifts~\cite{MFD96,MPD97}
(see however the optimistic estimates in~\cite{SFMP00}). As we have
mentioned at the beginning of the section, it is likely that the total
(galactic + extra-galactic) isotropic component of the GGB does not
exceed the total extra-galactic GGB; indeed, we will assume that the isotropic
portion of $\Omega_g(f)$, the only one that affects searches for the PGB,
follows the frequency distribution predicted by Eq.~(\ref{Sncg}), 
but reduced by a factor $\epsilon \le 1$. 
In the rest of the paper we will therefore assume
\be
\Omega_{g,is} = \epsilon \frac{4\pi^2}{3 H_0^2}\,f^3\,S(f)\,.
\label{omegag}
\ee
Note that the determination of the value of $\epsilon$ is a delicate matter.
At frequencies
above a few mHz, where the GWs from galactic sources can be subtracted, it is likely that
$\epsilon \simlt 0.3$, while below $\sim 1$ mHz, Ref.~\cite{GP97} suggests 
$\epsilon \approx 0.7$. However a more detailed analysis is needed in the future to
provide a better estimate of this value. In the following we optimistically set
$\epsilon = 0.1$, which can be regarded as a solid lower limit; the results that we
will present in the next Section can be easily rescaled 
as a function of $\epsilon$.

We consider now the critical frequency $f_g$ up to which radiation 
emitted by solar-mass binary systems (in the whole Universe) produce a GGB. For
$f\simgt f_g$ the observational window becomes
"transparent" to the primordial GW background. Following
the discussion at the beginning of the section, we estimate $f_g$ by using
a very simple argument based on first principles: if
the number of independent degrees of freedom of the data set -- the number of
data points, or, equivalently, the number of frequency bins -- is smaller than the
total number of independent parameters that describe the radiation,
then the superposition of monochromatic GWs must be considered as a stochastic background. 
If the opposite is true,
we have enough information to characterize, at least in principle, each individual
source, and the signal is a deterministic one. For the sake of simplicity -- although not
exactly true, see the discussion at the beginning of the section 
and~\cite{Cutler98} -- we assume that each binary system is characterized by one
parameter. 
The critical frequency $f_g$ is, therefore, formally determined by the 
condition that the (average) number of sources per frequency bin is less than one:
\be
\frac{dN(f)}{df}\,\Delta_{\rm b} f \simlt 1\,;
\label{fco}
\ee
here $dN/df$ is the number of binary sources, emitting at frequency $f$, per unit
frequency interval. Assuming that the merger rate is $R$, and 
in the relevant frequency range the binaries evolve only through radiation reaction, 
we have
\be
\frac{dN(f)}{df}\,\Delta_{\rm b} f = \frac{R}{T}\,\left(\frac{df}{dt}\right)^{-1}\,,
\label{dndf}
\ee
where $df/dt$ can be estimated using the Newtonian quadrupole formula: 
\be
\frac{df}{dt} = \frac{96}{5}\pi^{8/3}\Mc^{5/3} f^{11/3}\,,
\label{dfdt}
\ee
and $\Mc \equiv (m_1 m_2)^{3/5}/(m_1 + m_2)^{1/5}$ is the so-called 
{\it chirp mass} ($m_1$ and $m_2$ are the masses of the two orbiting stars). 
Substituting Eqs.~(\ref{dndf}), and~(\ref{dfdt})
into Eq.~(\ref{fco}), the frequency $f_g$ is easily determined:
\be
f_g \simeq
\left(\frac{5}{96}\right)^{3/11}\,\pi^{-8/11}\,
\Mc^{-5/11}\,\left(\frac{R}{T}\right)^{3/11}\,.
\label{fco1}
\ee

We are now interested in determining an upper limit to $f_g$, considering
binary populations from the whole Universe. For $f \simgt 10$ mHz, the main
contribution to the GGB is given by NS-NS binaries~\cite{HBW90}. Their merger rate 
is uncertain; current estimates yield a 
galactic merger rate in the range~\cite{Phinney,Narayan91,Kalogera00}:
\be
R_{\rm NS} \simeq 10^{-6} - 5\times 10^{-4}\,{\rm yr}^{-1}\,.
\label{ratens}
\ee
We can extrapolate this result to the entire
Universe by simply multiplying the galactic rate by the total number of galaxies $N_G$: 
\be
R \sim 10^6\,\left(\frac{R_{\rm NS}}{10^{-5}\,{\rm yr}^{-1}}\right)\,
\left(\frac{N_G}{10^{11}}\right)\,{\rm yr}^{-1}\,.
\label{Rtot}
\ee
By using this approach, we assume that
$R_{\rm NS}$ does not vary with the redshift, which is probably not true. However,
even if at high redshift $R_{\rm NS}$ is a factor 10 higher than in our galaxy,
the very weak dependence of $f_g \propto R^{3/11}$ on the merger rate, see
Eq.~(\ref{fco1}), ensures that this crude estimate is correct within a
factor $\simeq 2$. Assuming that the typical chirp mass of the binaries in the
population is $\bar{\Mc} = 1.2\,\Ms$, which corresponds to $m_1 = m_2 = 1.4\,\Ms$,
and the rate~(\ref{Rtot}), Eq.~(\ref{fco1}) yields:
\be
f_g \simeq 1.6\times 10^{-1} 
\left(\frac{R}{10^{6}\,{\rm yr}^{-1}}\right)^{3/11}\,
\left(\frac{T}{1\,{\rm yr}}\right)^{-3/11}\,
\left(\frac{\bar{\Mc}}{1.2\,\Ms}\right)^{-5/11}\,{\rm Hz}\,.
\label{fcot}
\ee
This simple analysis leads therefore to the conclusion that the window $f \simgt 0.1$ Hz
is likely free from stochastic backgrounds generated by astrophysical sources; 
radiation from NS-NS binaries is still present, but one can
detect each individual source, estimate its parameters, and remove from the data
stream the signals. In principle, the search for the PGB becomes limited only by
the instrumental noise.

\section{Sensitivity}
\label{sec:sens}

We can now proceed to discuss the sensitivity of LISA, and possible
follow-up missions, to search for the PGB. The major disadvantage of 
the presently designed LISA mission is the lack of 
two instruments with uncorrelated noise; in fact, each pair 
of the three co-located LISA interferometers share one
arm, and therefore common noise. One can, in principle, extract
two data streams with uncorrelated 
noise at all frequencies~\cite{Cutler98}. unfortunately they are equivalent 
to the outputs of a pair of detectors, one rotated by 45$^\circ$ with respect to the other,
and as we have stressed in Sec.~\ref{sec:overlap},
the response to a stochastic signal is then null. Nonetheless, LISA can play an important
role in searching for stochastic backgrounds, as it can possibly achieve
a sensitivity $h_{100}^2\Omega \sim 10^{-11} - 10^{-10}$
by exploiting some intrinsic properties of the signal and/or
the unique features of the instrument to operate in a configuration 
where the GW signal is (almost) absent. 

During the observation time, LISA changes 
orientation with period 1 yr. A background
which is anisotropic produces a periodic modulation in the auto-correlation function
that can stand above the noise~\cite{GP97}. Unfortunately, one
can not exploit this feature to detect a primordial signal 
that we expect to be intrinsically isotropic to a high degree; the only sizeable anisotropy
that one can foresee is a {\it dipolar} one, produced by the motion of our local system
(and therefore LISA) with respect to the cosmological rest frame with 
velocity $v_{\rm prop}/c \simeq 10^{-3}$. Unfortunately, an interferometer
has a quadrupolar antenna pattern, and the SNR produced by the quadrupole
component is reduced by a factor $(v_{\rm prop}/c)^2\sim 10^{-6}$. 
Nonetheless, the use of the signature induced by anisotropies could lead
to the detection of galactic generated backgrounds, as suggested in~\cite{GP97}.

Time delay interferometry with 
multiple readouts~\cite{TAE00} provides a way of suppressing by 
several orders of magnitude the GW contribution from the LISA output 
at frequencies below a few mHz, providing a shield to GW radiation. By exploiting
this feature, one can calibrate the noise-only response of the instrument, and search
for an excess power in the data stream that can be assigned to a signal of cosmic
origin. 

LISA can therefore carry out searches for stochastic backgrounds with a single
instrument at a very interesting sensitivity level. LISA is likely to detect the
galactic astrophysically-generated background, and could set important upper-limits
on the primordial contribution, that can rule out existing models. 
It also provides invaluable information regarding the astrophysically generated backgrounds
that might turn out to be crucial in designing future follow-up missions devoted to 
GW cosmology.
 
However, in order to reach the sensitivity $h_{100}^2 \Omega \sim 10^{-16}$ that we have
set as a goal, two separated interferometers are essential. It is important
to understand whether this target sensitivity is within the reach of near
future space technology, and to discuss possible fundamental limitations that might 
prevent us from
achieving a detection at the level $h_{100}^2 \Omega \sim 10^{-16}$. The science goals
are of such importance that some suggestions have 
been already put forward as to how to introduce modifications to the currently 
envisaged LISA configuration in order to accommodate a pair of independent 
instruments, possibly with the capability of shifting at-will
the centre of the sensitivity window from 
$\sim 10^{-3}$ Hz to $\sim 0.1$ Hz~\cite{Schutz00}. 

We start by considering a pair of identical LISA detectors; we therefore assume 
present, or near future, technology. Then, we discuss "second generation" LISA
detectors, specifically aimed at PGB searches.

\subsection{The sensitivity of two LISA interferometers}
\label{subs:sLISA}

We consider what the present LISA technology allow us to
achieve by cross-correlating the outputs of two identical interferometers
located at a distance $D$, cf. Eq.~(\ref{D}); this is equivalent to 
determining the minimum value of $\Om$ that one is able to detect. No unique answer can be given
to this question, as it depends, of course, on the 
frequency dependence of the true signal $\Omega(f)$. Lacking 
any solid theoretical prediction, 	
we choose a "maximum ignorance" approach, and assume that $\Omega(f)$ is constant over the
relevant frequency range. This hypothesis is not unreasonable, because 
the frequency band over which the SNR builds up is fairly small, due 
to $S(f)$ and $\gamma(f)$, cf. Eqs.~(\ref{Sni}),~(\ref{gamma1}), and 
Figs.~\ref{fig:noise} and~\ref{fig:gf}. By setting $\Omega(f) = {\rm const}$, and
$S_n^{(1)}(f) = S_n^{(2)}(f) = S_n(f)$, and solving Eq.~(\ref{snr_w}) for
$\Omega$, we can estimate the minimum detectable value of the energy density content of
the GW background%
\footnote{Notice that it is appropriate to use the weak (with respect to the noise) signal 
approximation, Eq.~(\ref{snr_w}), as we are aiming at the detection of the
weakest possible background, which is clearly dominated by the noise.}%
:
\be
\Omin = \frac{K}{T^{1/2}}\,\frac{\sqrt{50}\pi^2}{3H_0^2}\,
\left[\int_{0}^{\infty}
\frac{\gamma^2(f)}{f^6\,S_n^2(f)}\right]^{-1/2}\,;
\label{omegamin}
\ee
the constant $K$ is related to the false alarm probability
and the detection rate associated with the measurement of a background with energy
$\Omega^{({\rm min})}$. For a false alarm probability of $5\%$, and a
detection rate of $95\%$, we have $K\simeq 3.76$ (the sum of the false
alarm probability and the detection rate need not to be 1, and these two
quantities are totally subjective)~\cite{AR99}.
We therefore compute Eq.~(\ref{omegamin}), with noise spectral density and
overlap reduction function given by Eqs.~(\ref{Sni}) and~(\ref{gamma1}),
respectively. The results presented here are computed, for sake of simplicity,
for the case $\Delta \alpha_0 = 0$, and the integration is carried out over
the frequency band $10^{-4} - 10^{-2}$ Hz. %
\narrowtext\noindent%
\begin{figure}[htb!]
\begin{center}
{\epsfig{file=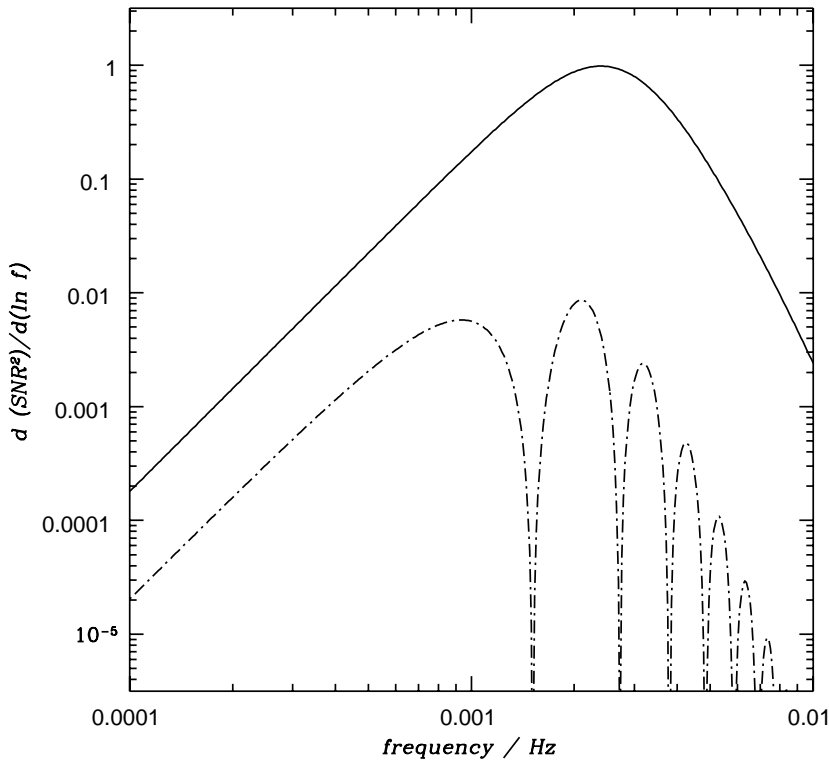,
angle=0,width=5.0in,bbllx=25pt,bblly=50pt,bburx=590pt,bbury=740pt
}}
\caption{\label{fig:dsnrdf} 
The contribution of different frequency regions to the total
signal-to-noise. The plot shows the fraction of SNR$^2$, Eq.~({\protect{\ref{snr_w}}}),
that is accumulated per unit logarithmic frequency interval, $d({\rm SNR}^2)/d(\ln f)$, 
as a function of the frequency $f$, for cross-correlations involving a pair of LISA
instruments. The vertical axis is normalized so that 
${\rm SNR}^2 = 1$ for co-located and co-aligned interferometers, with an
arm opening angle of $60^\circ$. The
solid and dotted-dashed lines refer to $D = 0$ and $D = 1$ AU, respectively. In
both cases $\Delta\alpha_0 = 0$, and the noise spectral density is given by 
Eq.~({\protect\ref{Sni}}).
}
\end{center}
\end{figure}
\narrowtext\noindent%

It is interesting to
analyze first how different frequency regions of the whole sensitivity window 
contribute to the total SNR. Fig.~\ref{fig:dsnrdf} shows the fraction of the total
signal-to-noise ratio that is accumulated per unit logarithmic frequency interval; it indicates clearly that the key frequency
band is $\approx 8\times 10^{-4}\,{\rm Hz} - 5\times 10^{-3}$ Hz.%
\narrowtext\noindent%
\begin{figure}[htb!]
\begin{center}
{\epsfig{file=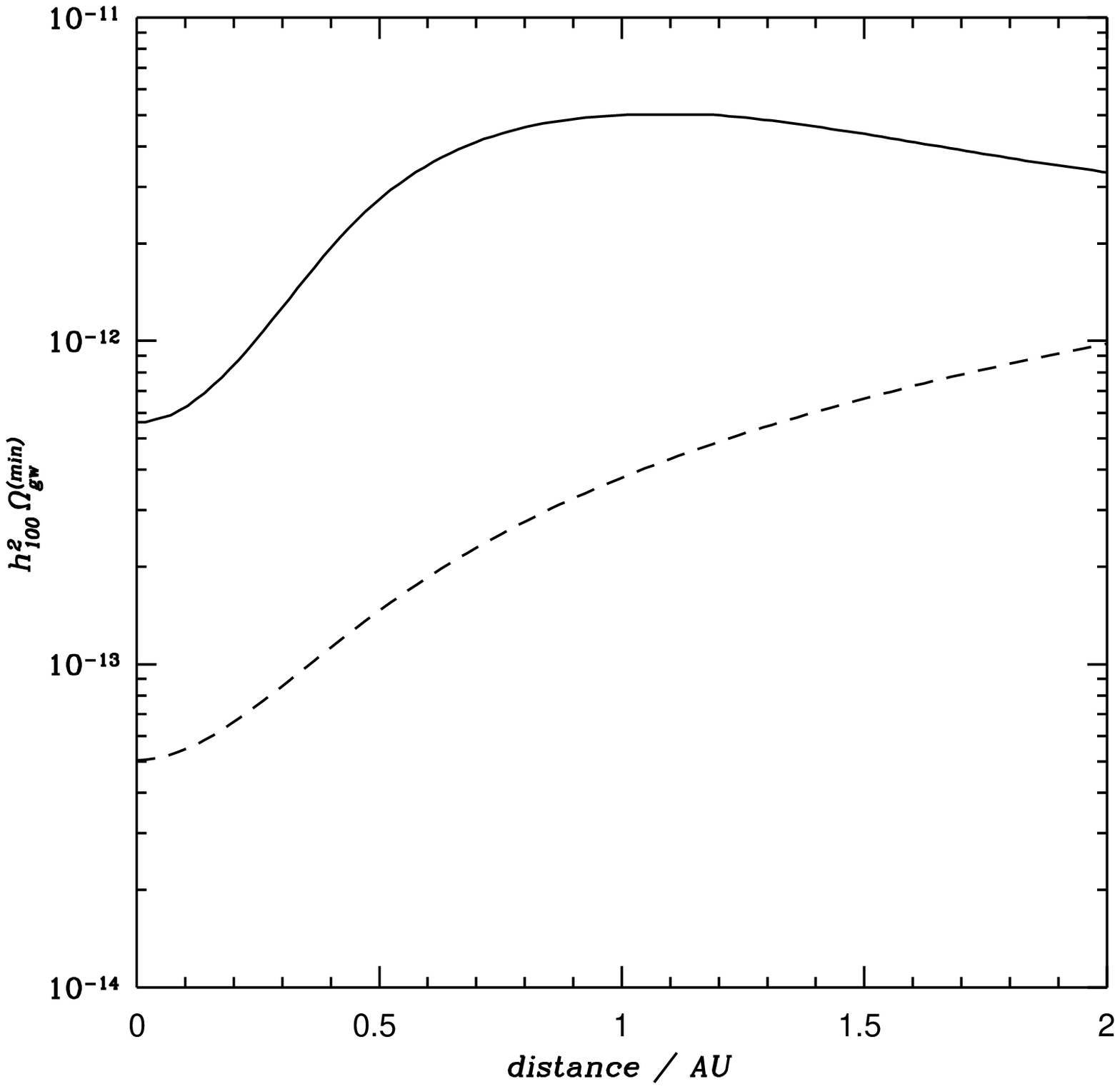,
angle=0,width=4.0in,bbllx=25pt,bblly=50pt,bburx=590pt,bbury=740pt
}}
\caption{\label{fig:ommin}
The minimum detectable value of the fractional energy 
density of a GW stochastic background. The plot shows 
$h_{100}^2\,\Omega^{({\rm min})}$ -- assumed to be flat, see 
Eq.~({\protect{\ref{omegamin}}}) -- for a cross-correlation experiment 
involving 
two identical LISA detectors, as a function of the instrument separation 
$D$ (in AU). The integration time is set to $T = 10^7$ sec.
The instrument noise spectral density and the overlap reduction 
function are given by Eqs.~({\protect{\ref{Sni}}}) 
and~({\protect{\ref{gamma1}}}), respectively; for the sake
of simplicity, we set $\Delta \alpha_0 = 0$ and the integration band 
corresponds to $10^{-4}\,{\rm Hz} - 10^{-2}$ Hz.
The dashed line refers to an experiment limited {\it only by instrumental
noise}, where the false alarm probability and the detection rate are 
$5\%$ and $95\%$, respectively; in this case $h_{100}^2\,\Omega^{({\rm min})}$
scales as $T^{-1/2}$. The solid line refers to an experiment limited
by {\it instrumental and confusion noise}, and shows the minimum detectable 
value of the energy spectrum of a primordial background 
$h_{100}^2\,\Omega_p^{({\rm min})}$. In this case, the spectrum of
the astrophysical generated background is computed according to 
Eq.~({\protect\ref{omegag}}), and we have optimistically set $\epsilon = 0.1$. 
The limit on $h_{100}^2\,\Omega_p^{({\rm min})}$ does not improve, in this
case, by increasing the integration time and/or lowering the instrumental noise.
We refer the reader to the text for further details.
}
\end{center}
\end{figure}
\narrowtext\noindent%

Fig.~\ref{fig:ommin} shows the sensitivity to a generic GW stochastic
background that one could {\it in principle} achieve
for a time of observation $T = 10^7$ sec. It is straightforward to rescale these
results for a different observation time, as SNR $\propto\, T^{1/2}$ and 
$\Omin\propto T^{-1/2}$, see Eqs.~(\ref{snr}) and~(\ref{omegamin}).
It is remarkable that LISA is capable 
of measurements in the range $5\times 10^{-14} \simlt \Omin \simlt 10^{-12}$, which is 
about three orders of magnitude better than can be achieved
(although in a different frequency regime) with Earth-based interferometers 
operating in the "advanced" configuration. The latter experimental set-up
requires considerable technological and possibly conceptual developments, and is expected 
to be implemented not before the end of the decade. The one-order-of-magnitude
range in $\Omin$ is due to the effect of the overlap reduction
function, where the lower limit is for co-located instruments, and the upper limit
for detectors at a distance D = 2 AU with $\Delta\alpha_0 = 0$, cf. also Fig.~\ref{fig:dsnrdf}.%
%
%
%
%

However, such instrumental sensitivity does not correspond to a comparable sensitivity 
to PGBs; in fact, the peak of the GGB spectrum, of the order $\sim 10^{-11}$ for
the isotropic component, is where the instrument is most
sensitive, cf. Fig.~\ref{fig:noise}. In order to quantify this effect, and to address 
the capabilities of space-based interferometers
for cosmology, we need a short digression.

Consider a generic two-component stochastic signal:
\be
\Omega(f) = \Omega_1(f) + \Omega_2(f)\,,
\label{Om2}
\ee
and assume that $\Omega_1(f)$ and $\Omega_2(f)$ share exactly
the same statistical properties (they are isotropic, stationary, Gaussian and unpolarized),
and therefore can not be distinguished from each other. In order to search
for $\Omega_1(f)$, one constructs the optimal filter
\be
\tilde Q_1 = \frac{\gamma(f) \Omega_1(f)}{f^3 R(f)}\,,
\label{Qp}
\ee
and correlates it against the data of two instruments, following the scheme described
in the Introduction, Sec.~\ref{subs:det}. The signal-to-noise ratio at the output of
the filtering process is
\ba
{\rm SNR}^2 & = & T\,\left(\frac{3 H_0^2}{10\pi^2}\right)^2\,
\frac{\left(\tilde Q_1,\left[\frac{\gamma(f) \Omega(f)}{f^3 R(f)}\right]\right)^2}
{\left(\tilde Q_1,\tilde Q_1\right)}
\nonumber\\
& = & T\,\left(\frac{3 H_0^2}{10\pi^2}\right)^2\,\left(\tilde Q_1, \tilde Q_1 \right)\,
\left\{ 1+
\frac{\left(\tilde Q_1,\tilde Q_2\right)}
{\left(\tilde Q_1,\tilde Q_1\right)}
\right\}^2\,.
\label{snr_pg}
\ea
The second term in brackets can be interpreted as the (undesired) 
residual correlation in the detection filter due to $\Omega_2(f)$, which can
not be eliminated. When it exceeds unity, the 
component $\Omega_1$ can not be detected, regardless of the instrumental
sensitivity; observations carried out with smaller noise disturbances, would simply 
increase both terms $(\tilde Q_1, \tilde Q_1)$ and $(\tilde Q_1,\tilde Q_2)$ 
by exactly the same amount, without
improving the chances of detecting $\Omega_1$. The
same holds for the integration time $T$: it has no effect at all on the capability
of discriminating the two components. Therefore, the minimum detectable value
of $\Omega_1(f)$ is set by the condition
\be
\left(\tilde Q_1, \tilde Q_1 \right) = \left(\tilde Q_1,\tilde Q_2\right)\,.
\label{lim}
\ee
Notice that the frequency dependence of the two components is very important: if 
$\Omega_1(f)$ and $\Omega_2(f)$ follow a similar frequency behavior,
the filter picks up more power from the "spurious" component $\Omega_2(f)$; if
they are drastically different, even if $\Omega_2(f)$ dominates $\Omega_1(f)$,
one could achieve a detection. %
\narrowtext\noindent%
\begin{figure}[htb!]
\begin{center}
{\epsfig{file=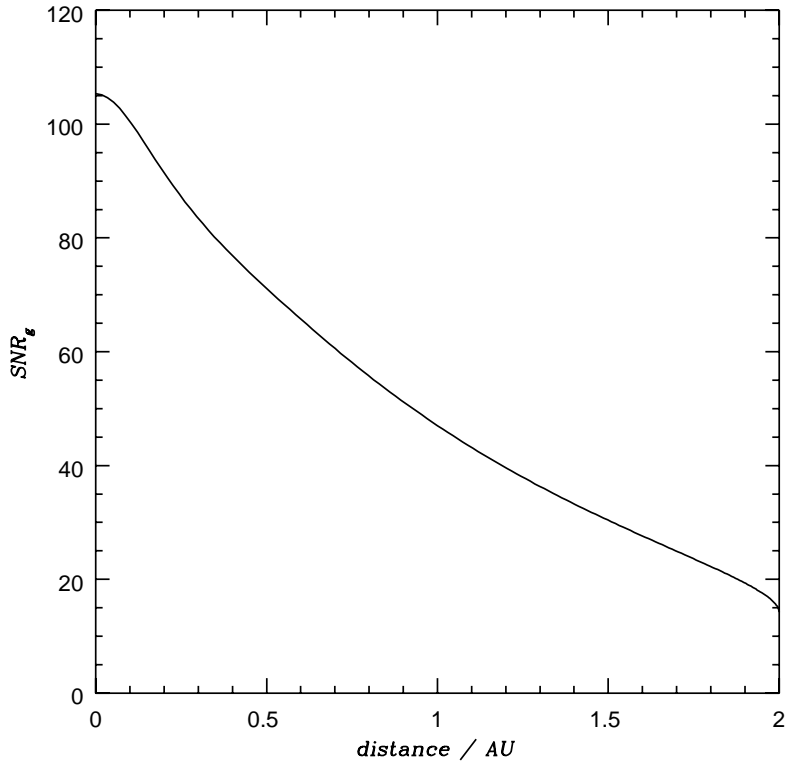,
angle=0,width=5.0in,bbllx=25pt,bblly=50pt,bburx=590pt,bbury=740pt
}}
\caption{\label{fig:snrc} The signal-to-noise ratio at which the astrophysically
generated background can be detected. The plot shows the SNR that can be achieved
in 4 months of integration time, correlating the data of a pair of LISA detectors, 
as a function of the distance $D$ (in AU). The Wiener filter is matched to the
signal described by Eq.~({\protect\ref{omegag}}).
}
\end{center}
\end{figure}
\narrowtext\noindent%

This example describes exactly the issue that we are considering in this section,
by simply identifying $\Omega_1$ with $\Omega_p$, and $\Omega_2$ 
with $\Omega_{g, is}$. 
The unresolved radiation from binary systems provides
therefore a {\it fundamental sensitivity limit} in searching for the primordial 
GW background.
We have computed this limit from Eq.~(\ref{lim}), in the case of an experiment carried
out with a pair of identical LISA's, assuming 
$\Omega_p$ constant, and $\Omega_{g,is}(f)$ given by Eq.~(\ref{omegag}).
The results
are summarized in Fig.~\ref{fig:ommin}; the key conclusion is that two LISA detectors will
be able to detect a PGB (with constant energy spectrum) only if 
$\Om_p \simgt 5\times 10^{-13}$. The big loss of SNR with respect to the case where
the experiment is limited only by instrumental noise is due to the fact that
the GGB is very strong in the mHz band,
and $\Omega_p(f)$ and $\Omega_g(f)$ have
a similar decreasing frequency behavior in the
frequency window where most of the SNR is accumulated: the residual correlation
at the filter output produced by the GGB is large. 

Given the results reported in Fig.~\ref{fig:ommin}, 
it is straightforward to conclude that the GGB is a guaranteed, strong GW
signal for space-based detectors. In fact, if one constructs a filter
matched to a GGB given by Eq.~(\ref{omegag}), and performs cross-correlations
between two identical LISA instruments which are characterized by the
noise curve~(\ref{Sni}), one can detect
such a signal with ${\rm SNR}_{\rm g} \sim 100$, for
a time of observation $T = 10^7$ sec; see Fig.~\ref{fig:snrc}; two to three
days of integration time are sufficient to reach
${\rm SNR}_{\rm g} \simeq 10$.
Two LISA-like detectors would therefore be extremely powerful
telescopes to launch deep surveys of populations of binary systems 
in our Universe with periods between a few hours and a few hundred seconds.

The bottom line of this analysis is therefore clear: 
{\it the fundamental limiting factor in searching for a primordial GW
background in the $10^{-5}\,{\rm Hz} - 10^{-2} \,{\rm Hz}$ frequency window is the 
stochastic radiation from unresolved binary systems}. Searches for PGBs with
$\Om_p \simlt 5\times 10^{-13}$ thus call for a change in the observational window. 
We briefly discuss this issue in the next Section.

\subsection{Towards testing slow-roll inflation}
\label{subs:sfut}

The mHz frequency window is unsuitable to reach the ambitious sensitivity level
$\Opmin \sim 10^{-16}$ predicted by slow roll inflation. One needs to
design an experiment with optimal sensitivity in a band free from generated backgrounds. 
Given our present astrophysical understanding, the
most promising region seems to be $\sim 0.1\,{\rm Hz} - 1\,{\rm Hz}$, which would be
optimally accessible through space-borne interferometers
with arms shorter than the LISA ones by a factor $\sim 100$. In fact,
the entire frequency window from $\sim 10^{-7}$ Hz to $\sim 0.1$ Hz is
contaminated by stochastic signals of astrophysical origin with $\Om \gg 10^{-16}$. 
A background generated by massive black hole binary systems
with fractional energy density $\Om \sim 10^{-15} - 10^{-14}$ is
present in the $\mu$Hz range~\cite{RR95}; however, our
ignorance concerning the formation rate of massive black holes and the
merger rate of massive black hole binary systems in the range $\sim 10^5\,\Ms - 10^9\,\Ms$
prevents us from giving a more solid estimate of the background generated by such objects. 
For $10^{-5}\,{\rm Hz} \simlt f \simlt 10^{-2}$ Hz,
unevolved binaries and WD-WD binary systems completely swamp the observational window;
see Sec.~\ref{sec:gb}. Above $\sim 10$ mHz the only residual 
sizeable contribution comes from NS-NS binary systems; as we have discussed in 
Sec.~\ref{sec:gb}, above $\sim 0.1$ Hz, the number of sources per frequency
bin becomes less than one, and the sky is ``transparent'' to 
a primordial signal. For rather long integration times $\approx 3$ yrs,
the rms instrumental noise level that is required to test the prediction from
slow-roll inflationary models is of the order $\sim 10^{-24}$
\be
\Opmin \simeq 8 \times 10^{-17} \,
\left(\frac{f}{0.1\,{\rm Hz}}\right)^{3/2}\,
\left(\frac{T}{10^8\,{\rm sec}}\right)^{-1/2}\,
\left[\frac{h_{\rm rms}}{10^{-24}}\right]^2\,.
\label{hrmshf}
\ee
Operating at considerably higher frequency than LISA, the two detectors would have
to be closely located, $D\simlt 10^{11}$ cm, in order to have optimal overlap reduction
function; however, this also increases potentially correlated noise sources.
Clearly the technological challenge to achieve the mentioned sensitivity is considerable; 
the main noise 
sources that would degrade the performance of such a detector are the shot 
noise, beam pointing fluctuations, and accuracy of the phase measurement 
technique.  This imposes stringent requirements on the power and frequency of 
the laser, as well as on the dimensions of the "optics" and on other components 
of the instrument. We would also like to stress that the value of $h_{\rm rms}$ 
that we have quoted in~(\ref{hrmshf}) is the {\it effective} noise fluctuation in the 
data stream, {\it after} the spectral
lines from individual NS-NS binaries, that are still copiously present in the
observation band, have been removed. How this can 
be effectively done and what instrument sensitivity is required is
still an open question that requires a careful analysis~\cite{Stebbins00}. 
The main result of
this crude analysis is that one could indeed reach the target sensitivity, and the 
fundamental limitations that make the mHz band unsuitable are removed. However,
several questions remain open for future analysis: the discussion of these issues 
goes far beyond the purpose of the present paper.

\section{Conclusions}
\label{sec:concl}

Gravitational wave experiments in the low-frequency window, 
together with high-frequency ground-based interferometers, are
expected to improve our picture of the very early 
Universe, and the understanding of the behavior of fundamental 
fields at high energy, by detecting, or setting stringent upper limits on
the primordial background of gravitational radiation. 

In this paper we have analyzed the sensitivity
of space-borne laser interferometers of the LISA class, and possible
succeeding missions. In order to set a reference frame for this discussion,
we have regarded the detection of a GW
background produced during the early Universe of energy $\Om_p \sim 10^{-16}$,
consistent with the prediction of standard slow-roll inflation, as the 
goal of GW cosmology.
We have assumed the operation of two space-detectors, in order
to achieve the best sensitivity and detection confidence, and we have shown that the 
technology available for LISA already ensures the detection of a
GW background
as weak as $\Om \approx 5\times 10^{-14}$. However, the strong stochastic
signal in the mHz band due to short-period solar-mass binary systems that
can not be resolved as individual sources prevents us 
from detecting a primordial background weaker than
$\Om_p \approx 5\times 10^{-13}$. Astrophysically generated
stochastic backgrounds therefore set a fundamental
limit in the mHz band that prevents us from achieving a sensitivity 
that goes beyond what is already guaranteed by the LISA technology. 
They also represent a guaranteed strong signal detectable at high signal-to-noise 
ratio, which enables the study of the distribution and merger rate of populations
of binary compact objects in the Universe. 

Dedicated missions with optimal sensitivity in the window
0.1 Hz -- 1 Hz appear, at present, the only viable option
in the search for very weak primordial backgrounds, and we have briefly discussed 
the technological challenges involved in probing slow-roll inflation. 
Our order-of-magnitude
analysis strengthens the hope that a sensitivity
level $\Om_p \sim 10^{-16}$ might be within the capability of future dedicated 
low-frequency detectors.

Our analysis clearly indicates the key issues that deserve further investigation:
a solid estimate of galactic and
extra-galactic GW backgrounds produced by astrophysical sources,
the investigation of the statistical issues that can lead to the discrimination
of the PGB from the GGB, and a more rigorous analysis of the technical 
and conceptual problems for low-frequency experiments dedicated to GW cosmology. 
On the observational side, the presently designed single-instrument
LISA mission is a fundamental step for the planning of more ambitious,
multi-detector experiments: we will be able to measure directly the
degree of anisotropy of the generated background, shedding light on the fundamental
limiting factor of mHz experiments. In fact, while the present paper
deals only with the detection of an isotropic stochastic signal,
the remarkable sensitivity of LISA offers
the chance of going far beyond: a detailed study of the anisotropy and angular
dependence of stochastic signals, both of astrophysical and primordial origin. 
Such an investigation is currently in progress, and will be reported in a separate publication~\cite{isotropy}.

\acknowledgments

We thank C.~Cutler and B.~Schutz for their help and encouragement
throughout this work. We especially thank P. Bender for sharing his latest 
thoughts on generated stochastic backgrounds, and for several comments
on a preliminary version of this paper. We would also like to thank K.~Danzmann
for illuminating discussions regarding LISA technology and the noise
sources at low frequencies. This work has also benefited from interactions with 
B.~Falkner and S.~Vitale.

\appendix

\section{}
\label{sec:app}

We give here the values of the numerical coefficients that enter 
expressions~(\ref{t1}),~(\ref{t2}) and~(\ref{t3})
of the overlap reduction function derived in Sec.~(\ref{subs:gammaL}):
\ba
A_0 & =& \frac{513}{4096}\,\,,A_1=\frac{135}{1024}\,\,,
A_2=\frac{243}{2048}\,\,,A_3 = -\frac{9}{1024}\,\,,
A_4=\frac{33}{4096}\,\,;
\\
B_0 & = & -\frac{27\sqrt{3}}{4096}\,\,,B_1= \frac{27\sqrt{3}}{1024}\,\,,
B_2= - \frac{81\sqrt{3}}{2048}\,\,,B_3 = \frac{27\sqrt{3}}{1024}\,\,,
B_4= -\frac{27\sqrt{3}}{4096}\,\,;
\\
C_0 & = &\frac{513}{8192}\,\,,C_1=\frac{171}{2048}\,\,,
C_2 =\frac{99}{4096}\,\,,C_3 = \frac{27}{2048}\,\,,
C_4=\frac{33}{8192}\,\,;
\\
D_0 & = & -\frac{27\sqrt{3}}{8192}\,\,,D_1 = -\frac{9\sqrt{3}}{2048}\,\,,
D_2 = \frac{63\sqrt{3}}{4096}\,\,,D_3= -\frac{9\sqrt{3}}{2048}\,\,,
D_4 = -\frac{27\sqrt{3}}{8192}\,\,;
\\
E_0 & = &\frac{513}{16384}\,\,,E_1=\frac{207}{4096}\,\,,
E_2 = \frac{339}{8192}\,\,, E_3 = \frac{63}{4096}\,\,,
E_4 = \frac{33}{16384}\,\,;
\\
F_0 & = & -\frac{27\sqrt{3}}{16384}\,\,,F_1= -\frac{45\sqrt{3}}{4096}\,\,,
F_2 = -\frac{177\sqrt{3}}{8192}\,\,,F_3 = -\frac{45\sqrt{3}}{4096}\,\,,
F_4 = -\frac{27\sqrt{3}}{16384}\,\,.
\ea

\end{document}